\documentclass[9pt,journal,compsoc]{IEEEtran}
\usepackage{ifpdf}
\ifpdf
  \pdfoutput=1\relax                   
  \usepackage{graphicx}                
  \DeclareGraphicsExtensions{.pdf,.png,.jpg,.jpeg} 
\else
  \usepackage{graphicx}                
  \DeclareGraphicsExtensions{.eps}     
\fi%

\graphicspath{{figures/}{pictures/}{images/}{./}{../figures}} 

\usepackage{hyperref}

\ifCLASSOPTIONcompsoc
  \usepackage[nocompress]{cite}
\else
  \usepackage{cite}
\fi
\usepackage{amsmath}
\ifCLASSOPTIONcompsoc
 \usepackage[caption=false,font=footnotesize,labelfont=sf,textfont=sf]{subfig}
\else
 \usepackage[caption=false,font=footnotesize]{subfig}
\fi
\begin{document}
%
\title{Large Scale Qualitative Evaluation of Generative Image Model Outputs}


%
%
%
%

\author{Yannick Assogba,
        Adam Pearce
        and Madison Elliott
\IEEEcompsocitemizethanks{\IEEEcompsocthanksitem Yannick Assogba, Adam Pearce and Madison Elliot are with Google Research.\protect\\
E-mail: \{yassogba,adampearce,madisone\}@google.com}
}

%
%

\markboth{}%
{Shell \MakeLowercase{\textit{et al.}}: Bare Demo of IEEEtran.cls for Computer Society Journals}
%


\IEEEtitleabstractindextext{%


\begin{abstract}
Evaluating generative image models remains a difficult problem. This is due to the high dimensionality of the outputs, the challenging task of representing but not replicating training data, and the lack of metrics that fully correspond to human perception and capture all the properties we want these models to exhibit. Therefore, qualitative evaluation of model outputs is an important part of model development and research publication practice. Quantitative evaluation is currently under-served by existing tools, which do not easily facilitate structured exploration of a large number of examples across the latent space of the model. To address this issue, we present Ravel, a visual analytics system that enables qualitative evaluation of model outputs on the order of hundreds of thousands of images. Ravel allows users to discover phenomena such as mode collapse, and find areas of training data that the model has failed to capture. It allows users to evaluate both quality and diversity of generated images in comparison to real images or to the output of another model that serves as a baseline. Our paper describes three case studies demonstrating the key insights made possible with Ravel, supported by a domain expert user study.
\end{abstract}

}

\maketitle

\IEEEdisplaynontitleabstractindextext

%
\IEEEpeerreviewmaketitle

\IEEEraisesectionheading{\section{Introduction}\label{sec:introduction}}

Generative image models are a class of neural network based models that aim to produce \textbf{novel}, \textbf{high-quality} and \textbf{diverse} images that faithfully model a target image distribution. A variety of architectures and training methods have been designed to learn such models, such as Generative Adversarial Networks (GANs) \cite{goodfellow_generative_2014}, Variational Auto-Encoders (VAEs) \cite{kingma_auto-encoding_2013}, Flow Based Models \cite{dinh_nice_2015} and Diffusion Models \cite{sohl-dickstein_deep_2015}.

Evaluating these models remains difficult \cite{odena_open_2019}. The high dimensionality of the output architectures used make likelihood estimates of model outputs difficult, and in some cases intractable. It has also been demonstrated that measures like average log likelihood do not always correlate with human perceptual judgments of sample quality \cite{theis_note_2016}. Additionally, while we want models to capture the target distribution well, we do not want them to produce images that are actually in the training set (an issue commonly referred to as memorization). 

A number of metrics have emerged in the literature around generative image models \cite{borjiProsConsGAN2018b} \cite{borji_pros_2021}, with Fréchet Inception Distance (FID) \cite{heusel_gans_2017} being the most popular. However, issues have been identified with FID, leading to the development of more granular metrics such as precision and recall \cite{sajjadi_assessing_2018} \cite{kynkaanniemi_improved_2019}.

Single-number metrics such as FID, while necessary for forward progress in the field, do not capture the full range of qualities desired of these models. Because of this, human visual inspection often plays a critical role in the evaluation and dissemination of advances in generative image modeling. However with existing evaluation tools, practitioners can typically only look at a small fraction of the output space of these models, on the order of 10s to 100s of images (e.g., \cite{brock_large_2019}, \cite{karras_style-based_2019}, \cite{kingma_glow_2018}, \cite{oord_neural_2018}, \cite{ho_cascaded_2021}). 

Our interviews with domain experts confirm that human evaluation is a critical part of practitioner workflows. Some experts rely on human evaluation with crowd-sourced evaluators, they however recognize that these are often expensive or time consuming and are thus left to the final stages of evaluation if done at all, leaving them to primarily rely on small scale qualitative evaluation during the model development process.

At the same time, experts in the field are concerned about cherry picking of results for publication but typically have no means to expansively explore model outputs in the rare occasions that these are published alongside academic manuscripts.

To address these needs, we built a system called \textit{Ravel}, which enables users to perform visual inspection of model outputs on scales up to three orders of magnitude greater that typical user workflows. We demonstrate usage of this system on datasets varying from 50k - 120k images. These dataset sizes are comparable to those used in standard \textit{quantitative evaluation} of generative image models.

Our primary contributions include:
\begin{itemize}
    \item A visual analytics system that supports \textit{multiple evaluation tasks} (e.g. evaluating quality \& diversity, discovering mode collapse or gaps in model output) for generative image models and is agnostic to model architecture and internals.
    \item \textit{Interactive exploration of large generative image model datasets}, facilitated by clustering and the use of fine grained visualization of cluster metrics to guide qualitative evaluation.
    \item A user interface that uses \textit{visual comparison} driven by semantically meaningful embedding spaces to support reasoning about differences between image distributions and generate hypotheses about model behaviour.
\end{itemize}

\section{Background}

\subsection{Generative Image Models}
The capabilities of generative image models have greatly increased over the last several years. Since the original GAN paper \cite{goodfellow_generative_2014} that broke the dam on modelling of faces, we now have systems like BigGAN \cite{brock_large_2019}, StyleGAN \cite{karras_style-based_2019}, GLOW \cite{kingma_glow_2018}, VQ-VAE \cite{oord_neural_2018}, CDM \cite{ho_cascaded_2021} and many others that present a wide variety of model architecture and training algorithms and are capable of producing very realistic images in a wide variety of domains.

\subsection{Quantitative Metrics for Evaluating Generative Image Models} \label{metricsformodels}
In this section, we outline the most commonly cited metrics in the research literature:

\begin{itemize}
    \item \textbf{Fréchet Inception Distance} \cite{heusel_gans_2017}: Uses a pre-trained InceptionV3 classifier \cite{szegedyRethinkingInceptionArchitecture2015a} to generate embeddings for both real and generated images, then uses a statistical measure to compare the distribution of embeddings from the two sources. FID is the most popular metric in the literature. It requires a large number of samples to produce an accurate estimate (generally at least 50k generated images), and cannot detect memorization of the training set. Karras et al \cite{karras2020analyzing} point out that the texture bias in ImageNet based CNNs like InceptionV3 \cite{geirhosImageNettrainedCNNsAre2019} imply that metrics derived from them will not capture all aspects of image quality.
    \item \textbf{Inception Score} \cite{salimans_improved_2016}: Uses a pre-trained inception classifier to measure, \textit{a)} how well each generated image matches a single ImageNet class, and \textit{b)} if the full the set of generated images has uniform coverage over all the ImageNet classes. Similar to FID, Inception Score requires a fairly large number of images and cannot be used to detect memorization. It also cannot measure intra-class diversity or detect mode collapse. See Barrat and Sharma \cite{barratt_note_2018} for a detailed discussion of issues with this metric. 
    \textbf{Both Inception Score and FID} are scalar scores designed to capture both image quality and diversity, and thus cannot reveal if the model is trading off one of these properties for the other to achieve a better score.
    \item \textbf{Precision and Recall Metrics}: To disentangle the measurement of image quality and diversity, Sajjadi et al \cite{sajjadi_assessing_2018} and Kynkäänniemi et al \cite{kynkaanniemi_improved_2019} propose precision and recall metrics to measure each independently. Broadly speaking, \textit{precision} corresponds to the sample quality, whereas \textit{recall} corresponds to the coverage of the sample distribution with respect to the target distribution - i.e. diversity.
\end{itemize}

These metrics are generally \textit{computed over the entire dataset}, and thus have low granularity. Even when they indicate that a model is better or worse, they don't specify where in the distribution of generated images improvements or regressions lie. Ravel increases the granularity of these metrics, providing a way to find specific clusters of images that score poorly on some metric

\subsection{Qualitative Evaluation / Visualization of Generative Image Model Outputs}

Due to the limitations of quantitative metrics discussed above, researchers also rely on visual inspection of model outputs to evaluate model performance. Visual inspection is typically performed during training, to monitor that the process has not immediately failed, as well as after training, to evaluate overall quality of the model. We validate and elaborate on this workflow and strategy with domain experts in Section \ref{current state of eval}. Although models often output 100s of thousands of images, our user study found that researchers are only able to inspect a small portion of images (in the 100s) during their analyses. 

Visually impressive samples are paramount for successfully publishing model advancements in scientific venues. Relevant papers in the field of generative image models typically include 10s-100s of images \cite{brock_large_2019} \cite{karras_style-based_2019} \cite{karras2020analyzing} \cite{menonPULSESelfSupervisedPhoto2020b}. This represents a very limited sample of the variety of images these models are typically trained to generate. Our user study also found that there is an assumption that authors "cherry-pick" the best images to include in their publications. Relatively few authors have published large datasets of un-cherry-picked output images alongside their publications. 

There are currently no purpose-built interfaces that make these convenient to browse or examine. For example, \cite{karras_style-based_2019} \cite{karras2020analyzing} each publish 100k output images to a publicly accessible Google drive. These images are organized into 1000 sub-folders to make the interface more usable given the large number of files. To view this output, users must either click through the folders individually, or bring their own interface to browse the images. 

\section{Related Work}

Borji \cite{borjiProsConsGAN2018b} \cite{borji_pros_2021} catalogues many of the metrics for automatic evaluation of generative models. Ravel utilizes the precision and recall metrics from \cite{sajjadi_assessing_2018} and \cite{kynkaanniemi_improved_2019}, but does \textbf{not} propose any new metrics. We thus situate this work in primarily relation to work on \textbf{qualitative evaluation} and \textbf{interfaces} to explore generative model output.

\textbf{Crowd-worker Evaluation}

Denton et al. \cite{dentonDeepGenerativeImage2015} use a small volunteer sample of human annotators to estimate quality by asking whether they can distinguish real from generated images. Zhou et al. \cite{zhouHYPEBenchmarkHuman} refine and scale this technique, asking crowd-sourced workers from Amazon's Mechanical Turk to make psychophysical judgments about real vs. generated images. While these methods are good at scaling up evaluation to larger dataset sizes, they are more time-consuming and expensive than manual inspection by researchers. Thus they are typically reserved for later stages of the evaluation pipeline. They also tend to focus on measures that can be evaluated on individual images (e.g. image quality), rather that corpus level properties (such as image diversity).  

By contrast, Ravel is designed to support \textit{researcher evaluation earlier in the development pipeline before the use of external raters,} and  allows researchers to evaluate \textit{both quality and diversity in the same interface}. It fits in between initial monitoring of training dynamics to ensure that the model is converging and larger scale human rater evaluation typically performed closer to model release. 

\textbf{Explaining Model Internals}

Bau et al. \cite{bauGANDissectionVisualizing2018} explore finding interpretable units (neurons) within GANs and visualizing the causal effects of ablating these neurons. Their method depends on having access to model internals and having a pre-trained object segmentation network to find objects within the scene to establish the casual relationship between neuron activation and network output. \cite{bauSeeingWhatGAN2019} Bau et al. use a pre-trained segmentation network to compare the distribution of objects found in generated images with those found in a set of real images. This provides a measure of diversity of the models outputs with respect to the objects that can be segmented by the pre-trained network. The authors also propose a method (Layer Inversion), to train networks that compute approximate \textit{inversions} of real images into the latent space of the model, to see what the network generates instead of the missing objects.

While these approaches are critical to better understanding of the internal mechanisms that drive model behavior, they are typically specific to a particular model architecture. Ravel treats models as black boxes and is thus \textit{agnostic to model architecture}. Ravel does use pre-trained networks to compute vector representations of images, but is less sensitive to the final task the pre-trained network is trained to perform. For example one could use the embeddings from the model under examination or any model that has learned semantically useful features such as InceptionV3.

\textbf{Online Exploration of Model Outputs}

White \cite{whiteSamplingGenerativeNetworks2016a} explores a variety of ways to sample images from latent space that enable repeatable visual comparisons between models. They introduce a number of visualizations designed to examine how models perform with respect to \textit{specific input images} that are used to test model behaviour. In a follow-up work White \& Loh \cite{whiteSpaceSheetNavigatingConceptual} introduce a novel visual interface based on a spreadsheet metaphor that allows users to use geometric operations in the latent space to interactively query these models and thus explore their output. 

While online methods enable users to explore specific hypotheses, they generally suffer from  supporting relatively small exploration spaces due to the slow generation of images. Ravel focuses on offline analysis of generated images, which allows examining much larger datasets and can thus complement online methods as means of generating hypotheses for further, more targeted investigation.

\textbf{Embedding Based Visualizations}

A number of works have visualized embedding spaces of large, high-dimensional datasets \cite{smilkovEmbeddingProjectorInteractive2016} \cite{hoFontMap} \cite{diagneCuratorTable} \cite{mcdonaldBirdSounds}.

Liu et al. \cite{liuLatentSpaceCartography2019c} present a visual analysis system which uses the latent space learned by the \textit{encoder} of a VAE to explore variation in an existing image dataset.  This task is conceptually similar to what we support in Ravel, however we do not attempt to learn a new latent space as we want to focus on the generator output and its latent space (rather than fixed input data). 

Ravel builds on these earlier works visualizing embedding spaces, and incorporates clustering to make navigating these spaces more tractable. We focus on dataset comparison as a way to ground exploration of the output space and cater to the needs of the image generation use case.

Xiang et al. \cite{xiang_interactive_2019} present a visual analytics system for correcting labelling errors in large datasets. They visualize t-SNE projections of image embeddings \textit{color coded by label to highlight data points that are mislabelled}. In contrast, Ravel is designed to support unconditional generation scenarios, where the generated (and ground truth images) \textit{have no labels}. Thus, rather than relying on labels for grounding, Ravel enables comparative evaluation of datasets to allow grounding evaluation of generated images in the distribution of real images.

\section{Approach \& System Design}

\begin{figure*}[ht]
  \centering
  \includegraphics[height=11cm]{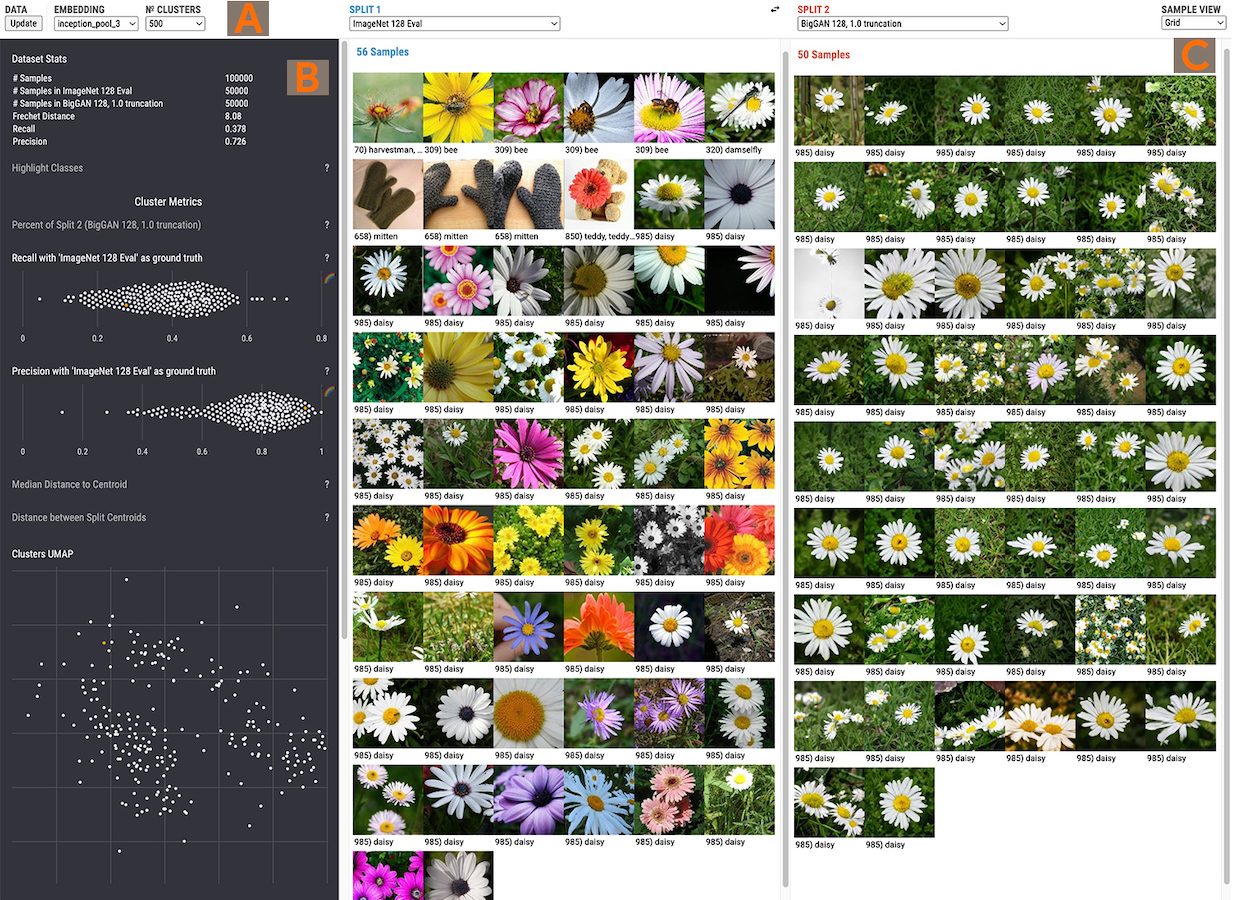}
  \caption{The Ravel interface primarily consists of: A) Dataset \& view options. B) Summary charts \& linked cluster plots. C) Side by side image grids for visual comparison of clusters. This view shows a cluster comparing real images on the left to generated images on the right.}
  \label{fig:teaser}
\end{figure*}

Ravel focuses on enabling visual inspection of model outputs in comparison with some baseline set of images, typically real 'ground truth' images. From our interviews (section \ref{current state of eval}), we know that practitioners perform visual analysis of image samples coming out of models, but typically on small numbers of images. \textit{Our aim is to support and scale up those workflows to allow more systematic visual inspection from 10s or 100s of images to 10,000s or 100,000s of images}. We leverage four key concepts to facilitate this comparison across a large set of images:

\begin{enumerate}
    \item \textbf{Image Embeddings}: Neural image embeddings provide a semantically rich vector space for images that allow computing similarity scores between pairs of images \cite{zhang_unreasonable_2018}. They have become the standard in evaluation pipelines for generative image models (section \ref{metrics}) and \textbf{we want to leverage the familiarity researchers have with embedding based metrics} in Ravel. The ability to compute semantic similarity between images allows us to group similar generated images together \textbf{and} allow comparison of those images to similar ground truth images. By default we compute the standard InceptionV3 embeddings used in metrics like FID, but can also add embeddings from other models including pre-trained models or from the model under examination if those are available.
    \item \textbf{Clustering Images}: In order to enable visual exploration of hundreds of thousands of images we need to group them into a smaller number of meaningful groups. We use unsupervised clustering of images to reduce the number of \textit{top-level items} the user has to consider from e.g. 100k images to 1000, 500 or even 250 clusters. We use k-means clustering and provide more details about this in the \textit{pipeline details} section below
    \item \textbf{Cluster Metrics}: Once we have reduced the number of top-level elements to a manageable number we wish to provide hints to the user as to \textbf{which clusters might be most interesting to explore first}, as well as \textbf{scaffold a repeatable workflow that can guide exploration}. We compute metrics over each cluster that enable sorting clusters into predictable order. Many of the metrics we compute are designed to surface differences between the generated data and the baseline data in a cluster, with others show general properties of the cluster itself. We provide more details about the metrics we compute in the \textit{pipeline details} section below. By sorting clusters by these metrics we allow discovery of outlier clusters as well as analyzing properties of different parts of the generative image space (e.g. seeing what kinds of output images have low precision)
    \item \textbf{Interactive Visualization}: Finally we provide responsive interactive visualization of images, clusters and associated cluster metrics. Because we compute all the data offline, the user interface is quite responsive and enable fast and free exploration of a large number of data-points.
\end{enumerate}

\subsection{Pipeline Details}

\subsubsection{Clustering}

Ravel clusters the image embeddings using the implementation of k-means clustering from scikit-learn. k-means was an attractive clustering algorithm to start with because it is a well known algorithm that has a single easily understood hyper-parameter, namely the number of clusters. This allows the user to \textit{directly specify values for this hyper-parameter that they believe make sense for their dataset}. In addition to this, compared to other algorithms with similar hyper parameter options such as spectral clustering, k-means scales well with the number of examples and number of clusters selected \cite{scikit_learn_docs_clustering}. k-means also tends to produce fairly evenly sized clusters, which is helpful for displaying their contents in a uniform way.

All clustering methods suffer from issues of imperfect cluster assignment, particularly at the boundaries of clusters. However, Ravel mitigates this issue somewhat by visualizing clusters according to their positions in latent space. This allows users to quickly discover and examine clusters that are near to each other and ascertain whether they are truly a single semantic cluster. For more details see the \textit{dimensionality reduction} section (\ref{dimensionality}) below.

\subsubsection{Cluster Metrics} \label{metrics}

We compute a number of metrics for each cluster. Having multiple metrics provides different lenses through which to consider the clusters. For example one might be interested in where recall or precision are low, or alternatively where clusters are tightly packed or more spread out.

Clusters may have images from either the generated data or the baseline/ground truth, in the UI we refer to these data sources as \textit{splits} and typically display the \textit{baseline data on the left}, though the user can change this interactively. Many metrics are geared toward exposing differences between the two splits that are currently being explored.

\begin{itemize}
    \item \textbf{Percent of Split 2:} The percentage of images in the cluster that come from the split displayed on the right (usually the generated data).
    \item \textbf{Recall:} Recall is a metric that describes what fraction of samples in the \textit{ground truth} split have support in the alternate split. We compute recall as defined in \cite{kynkaanniemi_improved_2019} but aggregate the per-image values for each cluster.  When comparing generated images to real ones a high recall implies the model is producing images as diverse as the real data.
    \item \textbf{Precision:} Precision is an metric that describes what fraction of samples in the generated data have support in the ground truth. As above we compute precision as defined in \cite{kynkaanniemi_improved_2019} but aggregate the per-image values for each cluster. When comparing generated images to real ones a high precision implies better image quality/realism.
    \item \textbf{Distance between split centroids:} Measures the distance between centroids of the samples in a cluster that belong to each split. Larger values suggest that the data from the two splits are more visually different while smaller values suggest that the left and right split are more visually similar.
    \item \textbf{Median distance to centroid:} Median distance of samples in the cluster to the centroid of that cluster. Smaller values here imply more compact clusters with more similar samples, higher values suggest the cluster has a greater variety of images.
\end{itemize}

\subsubsection{Dimensionality Reduction} \label{dimensionality}

In addition to visualizing clusters by the metrics described above, we also visualize the positions of cluster centroids in the embedding space itself. Since these are high dimensional embeddings spaces we need to use dimensionality reduction in order to visualize them in 2D. We use the UMAP algorithm \cite{mcinnes2018umap-software} to do this projection. Other options for dimensionality reduction include PCA \cite{pca1901} or t-SNE \cite{van2008visualizing}. Compared to t-SNE, UMAP has been reported to preserve more of the global structure of the original space and is significantly more computationally efficient both as the number of dimensions increases and as the size of the dataset grows  \cite{mcinnes2018umap-software} \cite{coenen_understanding_2019}. The InceptionV3 embedding we use is 2048 dimensional embedding, making a computationally efficient method particularly attractive. While not as directly interpretable as linear methods like PCA, UMAP is better able to capture some of the complex non-linear relationships between images encoded in the embedding space.

\subsection{User Interface}
The user interface is a browser-based application. We describe the design of the user interface and further discuss the utility of these affordances in the case study section.

\autoref{fig:teaser} shows the Ravel interface, divided into 3 main sections:

\subsubsection{(A): Dataset and View Controls}

The user can select which embedding to use, the number of clusters and which split to show on the left or right.

\subsubsection{(B): Charts}

There are three main kinds of data display in the charts section of the UI. 

\textbf{Summary Statistics}

First is a static table showing summary statistics and metrics for the dataset as a whole. These include things like the number of samples in the whole dataset, the number of samples in each split as well as dataset level metrics such as Fréchet distance\footnote{When using the Inceptionv3 embedding this is Fréchet Inception Distance (FID)}, recall and precision \cite{kynkaanniemi_improved_2019}.

\textbf{Cluster Metric Plots}

Below the summary charts are a series of beeswarm plots for the per-cluster metrics that were computed (see \textit{Cluster Metrics} section above for details). Each beeswarm shows \textbf{each cluster as a dot} and plots the distribution of clusters over that metric. These charts are interactive, individual dots can be selected to show the images from that cluster in the sample viewer. 

\begin{figure}[ht]
    \includegraphics[width=\columnwidth]{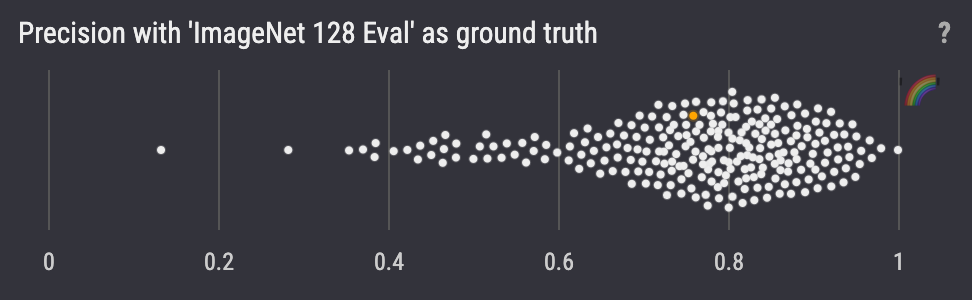}
    \caption{Beeswarm plot showing distribution of cluster precision scores. Each dot is a cluster which the currently selected dot shown in orange. A description of the metric can be accessed by clicking on the ? icon.}
    \label{fig:beeswarm_single_plain}
\end{figure}

\begin{figure}[ht]
    \includegraphics[width=\columnwidth]{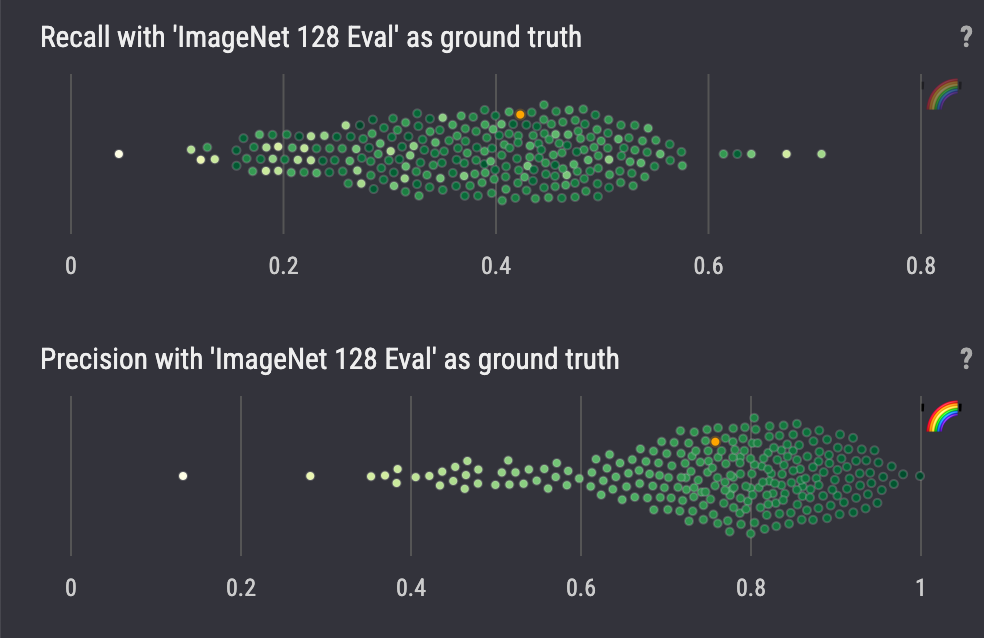}
    \caption{Recall and Precision beeswarm plots colored by \textbf{precision} score. High recall, low precision clusters can be identified using the position encoding in the recall plot and the color encoding applied from the precision plot. To save space in the display we do not display a legend, as the relationship between high-low values and hue is directly visible in the chart whose rainbow icon was toggled and exact values are of less importance than relative comparisons}
    \label{fig:beeswarm_rainbow}
\end{figure}

A color encoding can also be applied by clicking on the rainbow icon toggle next to that plot. This color encoding is applied across all charts allowing comparison of one metric to another (\autoref{fig:beeswarm_rainbow}). 

\textbf{UMAP Plots}

\begin{figure}[ht]
    \centering
    \includegraphics[height=11.0cm]{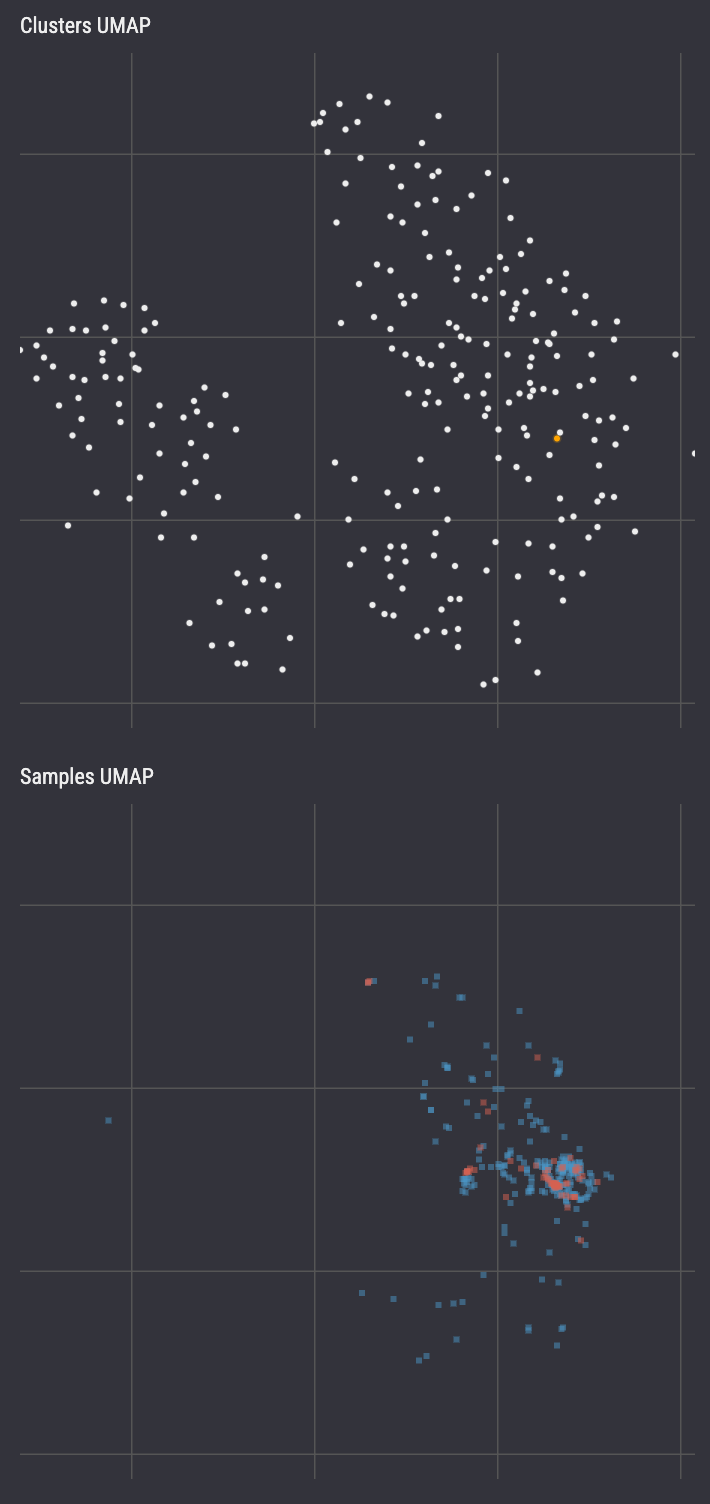}
    \caption{Top: UMAP projection of \textbf{cluster centroids} in embedding space. Bottom: UMAP projection of currently selected cluster, real samples are shown in blue and generated samples are show in red}
    \label{fig:umap_plots}
\end{figure}

Below the beeswarm plots are two plots showing 2D UMAP projections of: a) all the clusters and b) images from the \textit{currently selected cluster} (See \autoref{fig:umap_plots}). The Cluster UMAP view in particular allows browsing clusters by visual similarity rather than by the metric scores. Previews of individual images are shown on mouseover of points on the Samples UMAP plot 

\textbf{Highlight Classes}

In cases where the dataset does contain class labels, Ravel will display a search interface that allows users to search for and select any number of matching classes in the dataset.

\begin{figure}[h]
    \centering
    \includegraphics[height=8.5cm]{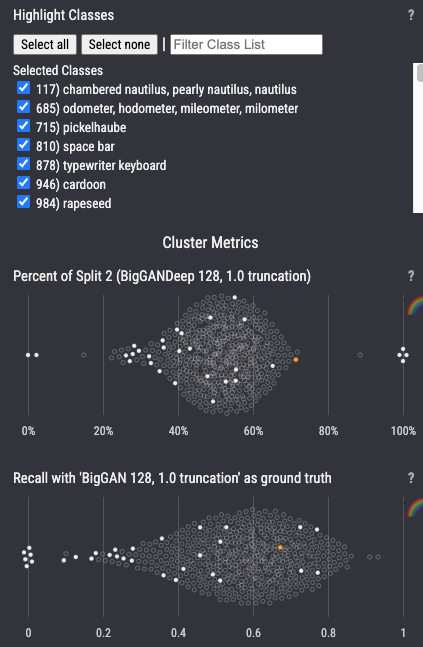}
    \caption{ Clusters containing at least one image with the selected classes are highlighted in all cluster plots.}
    \label{fig:biggan_highlight_classes}
\end{figure}

If one or more classes is selected Ravel will dim clusters that do not contain any images from those classes in the cluster plots (\ref{fig:biggan_highlight_classes}). 

All the cluster plot interactions are linked, enabling users to see where a cluster falls in multiple metrics or in embedding space. 

Any chart can be collapsed by by clicking on its title, this allows making more room for the charts the user finds most useful for their analysis.

\subsubsection{Sample Viewer} \label{sample viewer}

To the right of the charts is the sample viewer. This consists of a pair of scrollable views displaying image thumbnails from the selected cluster divided by split. If there are classes/labels associated with the images they are displayed below the thumbnail and images from the same class are grouped together, otherwise they are displayed in the order they were processed by the pipeline (effectively random). Users can click on images to get a larger view of the image and see any additional metadata.

Image thumbnails are displayed at a maximum size of 150x150px and a minimum size of 100x100px depending on how large the browser window is. On a 3840 x 2160 resolution display\footnote{This resolution is the default \textit{4K} resolution which is widely available. It is also close to the native resolution of a 2021 14 inch MacBook Pro (3024 x 1964)} one can typically see 72 images per split for a total of 144 images in a single screenful. However one clear limitation is that if there are more images than this in a cluster a user cannot see them all at once and does have to scroll through to get a better understanding of the cluster. One way to mitigate this is to use choose a higher number of clusters to view. Which results in smaller, more granular clusters, at the cost of having more to explore. 

\section{Case Studies}

In this section, we illustrate how the design features of Ravel support user exploration in a series of case studies. These case studies highlight discoveries the authors made when using the tool. Our 3 use cases are: 1) Unconditional image generation in a single domain (faces), 2) Class conditioned generation on ImageNet data and 3) Analyzing downstream use of a generative model for another task - in this case, super-resolution.

While Ravel supports the use of multiple embeddings for clustering and exploration, in this paper we focus primarily on \textit{the same Inceptionv3 embeddings used in FID score and other metrics}, as they have become a de-facto standard in the generative model space and are publicly available.

\subsubsection{Unconditional Image Generation}

Unconditional generation refers to situations where a trained model is used to generate images with no 'conditioning' other than an input vector (often referred to as a 'noise' or 'Z' vector) drawn from a random distribution. In this case study, we generate 60k images from an implementation of the StyleGAN2 architecture \footnote{We want to make clear that we are using an independently trained StyleGAN2 and not the StyleGAN2 weights released by the original authors} \cite{karras2020analyzing} trained on the FFHQ dataset. The images are generated using Z vectors drawn from a random distribution with a mean of zero and a standard deviation of one. We load those images, along with 67,542 images from the training set for a total of 127,542 images.

\emph{\textbf{Are there images that this model generates that are not part of the input distribution?}}

\begin{figure}[t]
 \centering
 \includegraphics[height=8.0cm]{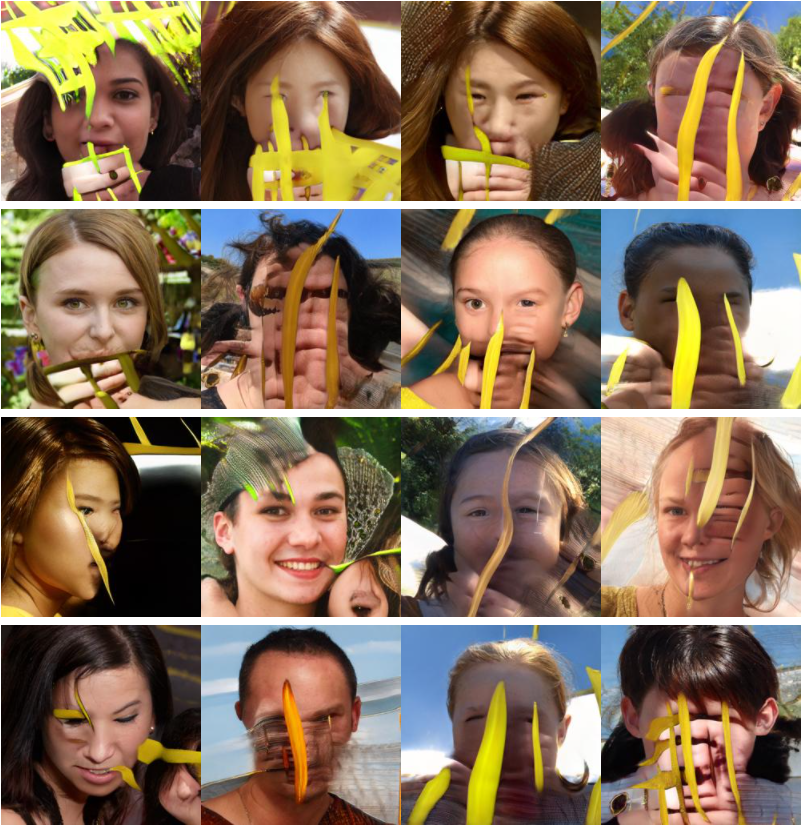}
 \caption{The StyleGAN2-based model struggles to model images of faces with facepaint and other colorful accessories occluding the face, it instead produces these artefacts.}
 \label{fig:stylegan2-facepaint-artefacts}
\end{figure}

Our analyst opens the Ravel interface in her browser and sets the number of clusters to 250, with images from the training data on the left and generated images on the right. Knowing that low precision generally implies less realistic images, she begins by exploring clusters on the low end of the precision chart and notes that the clusters with the 10 lowest values each contain generated images with visually obvious defects. Seeing these images contrasted with ground truth data gives her a sense of what the model is struggling to capture in each category. For example she notes that the model has a difficult time modeling complex backgrounds, portraits where there are more than one face, or portraits with occluding objects like hands or microphones. Another problem that stands out to her is the difficulty the model has generating faces with face-paint as in \autoref{fig:stylegan2-facepaint-artefacts}. Seeing these corrupted images juxtaposed with real images from the training data allows her to hypothesize why the model struggles with this, in particularly she observes that these types of images are relatively rare in the training data.

In examining 10 clusters, she has quickly scanned through approximately 4234 generated images and 2718 real images.

\subsubsection{Class Conditioned Image Generation}

Class conditioned generation refers to models that have been trained to generate output images for a fixed set of distinct classes. Here, we look at directly comparing the output of two models trained on the same data and task: namely, generating images from 1000 classes of the ImageNet dataset \cite{ILSVRC15} at a resolution of 128x128px. We use model implementations from Lucic et al's "Are GANs Created Equal? A Large-Scale Study" \cite{lucic_are_2018}. 

We set up a Ravel instance with 50k images generated by BigGAN 128 and BigGAN-deep 128 \cite{brock_large_2019}, and 50k images from the validation set for ImageNet\footnote{Data retrived from https://www.tensorflow.org/datasets/
catalog/imagenet2012}. Input vectors for each model are drawn from a random normal distribution, as described in \cite{lucic_are_2018}, and no truncation is applied to the input vectors.

Using these models demonstrates a workflow where researchers are trying to determine how one variant of a model, or an improved model architecture, behaves differently from some baseline model. This comparison is a common workflow in machine learning, typically achieved by reporting performance on key metrics such as FID. We demonstrate how Ravel can complement traditional metrics to find specific differences in model behaviour.

\begin{figure}[ht]
    \centering
    \includegraphics[height=6.5cm]{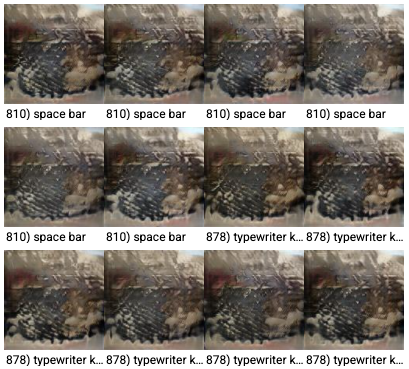}
    \caption{Mode collapse of the "space bar" and "typewriter keyboard" classes in the BigGAN-deep model. All 100 images in this cluster look identical as both classes have collapsed onto the same output.}
    \label{fig:BigGAN-deep_spacebar_typewriter}
\end{figure}

\begin{figure}[ht]
    \centering
    \includegraphics[height=6.5cm]{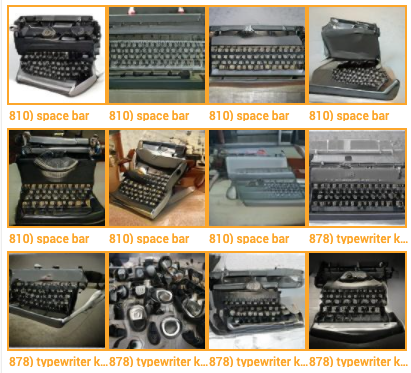}
    \caption{Output of the "space bar" and "typewriter keyboard" classes in BigGAN show much more variety. This model does not exhibit mode collapse for these classes}
    \label{fig:biggan_spacebar_typewriter}
\end{figure}

\emph{\textbf{How does BigGAN compare to BigGAN-deep in terms of diversity of output}}

Our analyst opens Ravel and sets the number of clusters to 500, with the left split showing output from BigGAN, and the right split showing output from BigGAN-deep. 

It immediately jumps out to him that there are a number of clusters that have scores of 0 in the recall chart. He clicks on one of them and discovers it only contains images from BigGAN-deep. All 100 images from this model are from two classes, appear virtually identical, and are of low visual quality (see \autoref{fig:BigGAN-deep_spacebar_typewriter}). This is a classic case of mode collapse \cite{goodfellowNIPS2016Tutorial2017}; the model was unable to learn the true distribution for these classes and has 'collapsed' all output for these classes to a single image. In this case, both classes have been collapsed into the same output image. The analyst clicks on other clusters with zero recall and finds similar mode collapse in BigGAN-deep for the following classes: \textit{"space bar", "typewriter keyboard", "pickelhaube", "rapeseed", "cardoon", "chambered nautilus, pearly nautilus, nautilus", and "odometer"}.

Using the "highlight classes" feature of Ravel \ref{fig:biggan_highlight_classes}, our analyst is able to find all clusters that contain images from these classes, and confirms that the BigGAN model \textit{does not} show mode collapse for these classes (\autoref{fig:biggan_spacebar_typewriter}). 

\textbf{\textit{What about classes where both models do okay at generation (i.e. neither model exhibits a pathological failure like mode collapse)?}} \label{sea urchins}

\begin{figure}[ht]
    \centering
    \includegraphics[width=\linewidth]{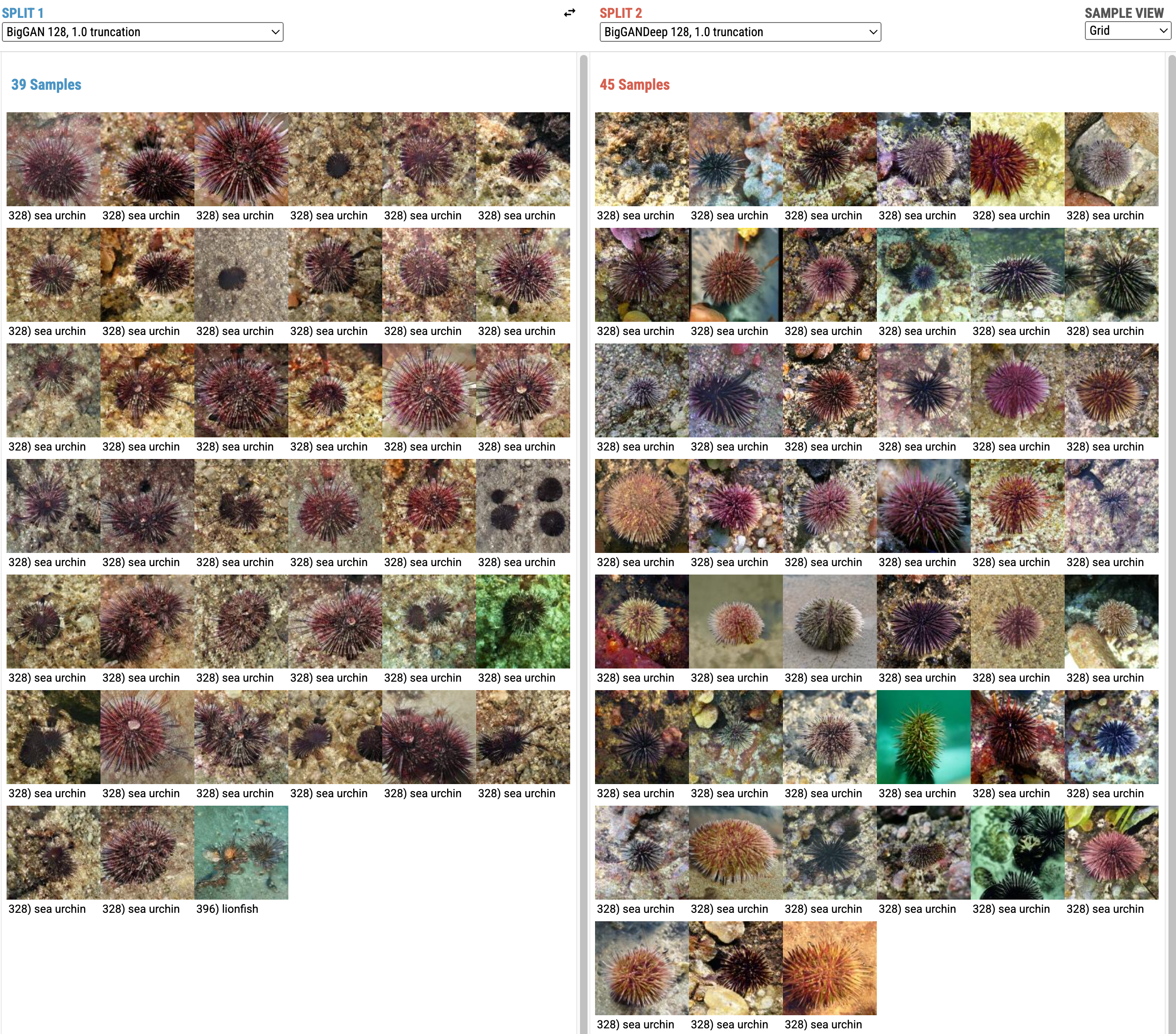}
    \caption{A cluster of sea urchin samples from two generators, both produce high quality images, but the generator on the right produces more diverse output. The images on the right show a greater variety of colors, textures and poses.}
    \label{fig:biggan_sea_urchin}
\end{figure}

Our analyst now chooses to increase the granularity of clusters by setting the number of clusters to 1000. He turns on the \textit{color by} option of the \textbf{precision chart}, and then looks at the \textbf{recall chart} to find clusters with high recall but relatively low precision (though not as low as in the previous section). Looking at higher recall clusters allows finding ones that have at least some overlap in their distribution, while keeping an eye on precision suggests differences in quality. These appear as light yellow dots on the right side of the recall chart. He randomly selects one, and discovers a cluster of sea urchins. Both generators produce high quality outputs, however visual inspection shows that BigGAN-deep produces more diverse output (\autoref{fig:biggan_sea_urchin}). 

\begin{figure}[ht]
    \centering
    \includegraphics[width=\linewidth]{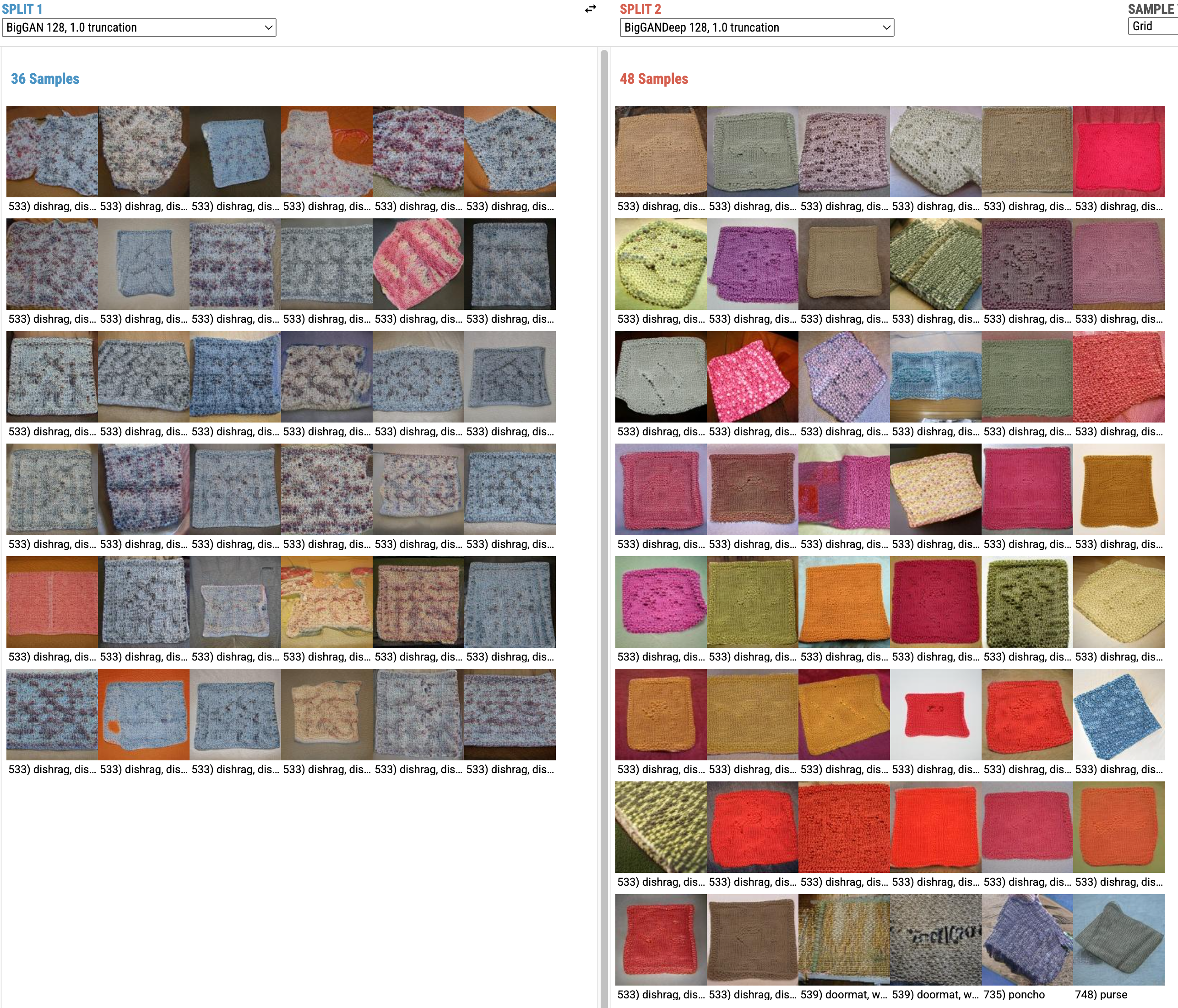}
    \caption{A cluster of 'dishrag' samples from BigGAN on the left and BigGAN-deep on the right, both produce somewhat realistic dishrags, but the generator on the right produces images with more diverse textures and colors.}
    \label{fig:biggan_dishrag}
\end{figure}

\begin{figure}[ht]
    \centering
    \includegraphics[width=\linewidth]{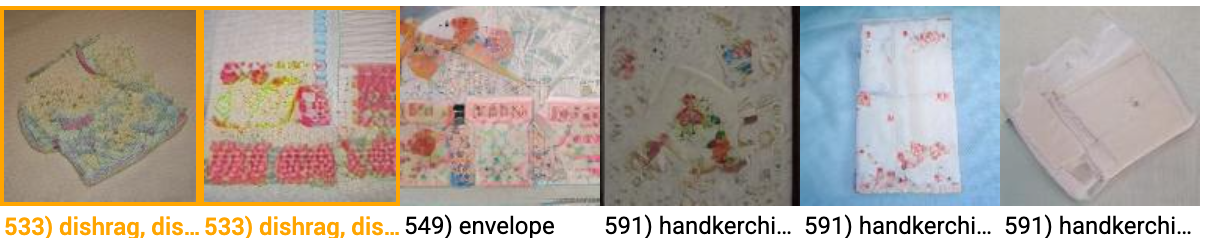}
    \caption{BigGAN cluster with a few dishrags but many colorful handkerchiefs that have similar patterns to those seen in the main dishrag cluster. This suggests some leakage in visual representation between the two concepts.}
    \label{fig:biggan_dishrag_handkercheif}
\end{figure}

\begin{figure}[ht]
    \centering
    \includegraphics[width=\linewidth]{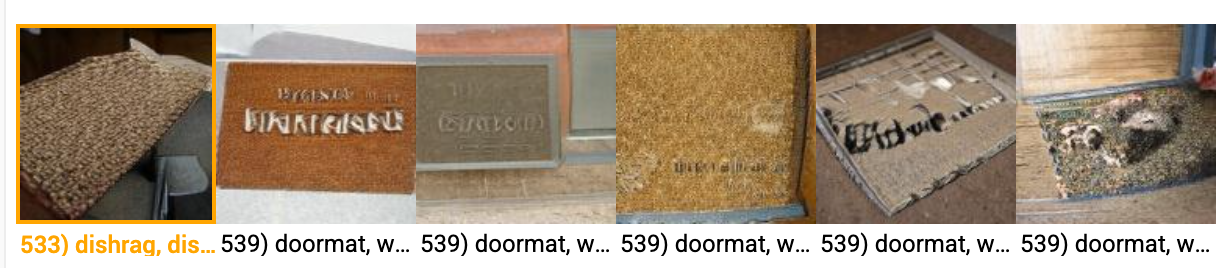}
    \caption{BigGAN-deep cluster with a few dishrags but many doormats, this suggests a hypothesis for why many of the BIgGAN-deep dishrags have a very rough texture.}
    \label{fig:biggandeep_dishrag_doormat}
\end{figure}

A benefit of clustering by visual similarity, rather than grouping by label, is making it easier to discover overlap or leakage of visual semantics between classes. Using the same method as above, our analyst selects another cluster that consists of relatively realistic images of dishrags (\autoref{fig:biggan_dishrag}). He then highlights all clusters containing dishrags. He can now individually examine all these clusters. His findings, shown in (\autoref{fig:biggan_dishrag_handkercheif} and \autoref{fig:biggandeep_dishrag_doormat}), suggests that other classes, namely handkerchief' and 'doormat', are 'leaking' into how each generator models dishrags.

\subsubsection{Downstream Applications of Generative Models: Super-Resolution}

In our final case study we consider the PULSE (Photo Upsampling via Latent Space Exploration) algorithm \cite{menonPULSESelfSupervisedPhoto2020b}, a super-resolution algorithm that searches the manifold of a generative image model (in this case StyleGAN2) to create plausible upsampled images corresponding to low resolution input images. The post-publication release of this model received sharp critique, as users quickly discovered weaknesses in the models ability to upsample images of people with non-caucasian features \cite{LessonsPULSEModel2020}. We use this example to illustrate how broader exploration of output manifolds could help researchers and practitioners who are using these kinds models in downstream applications to understand their weaknesses and mitigate risks before release.

In this scenario, we use the original CelebA-HQ dataset from Karras, et al. \cite{karras_style-based_2019}. This dataset consists 30,000 facial portraits of celebrities at 1024x1024px. We take CelebA-HQ images at 32x32px and upsample them back to 1020x1024px using the PULSE algorithm\footnote{This scale factor (24x) is within the range of scale factors (8x-64x) the original authors used in evaluation}. We invoke PULSE with the default settings provided by the authors in their model release \footnote{https://github.com/adamian98/pulse}, however we increased the number of steps we run the algorithm from 100 to 200 steps to increase the chances of PULSE producing a result for a given input.\footnote{We did this because PULSE will not always produce a result for an input image, and we found increasing the number of steps allowed substantially more input images to produce some result.}

We then compare the original 1024x1024px images and the PULSE up-sampled ones using Ravel. In total, we have 30000 original CelebA-HQ images and 22092 images output from PULSE. 7908 images failed to produce any output after 200 steps of the PULSE algorithm.

\emph{\textbf{What kinds of images does PULSE struggle to produce any output for?}}

\begin{figure}
    \centering
    \includegraphics[height=6.2cm]{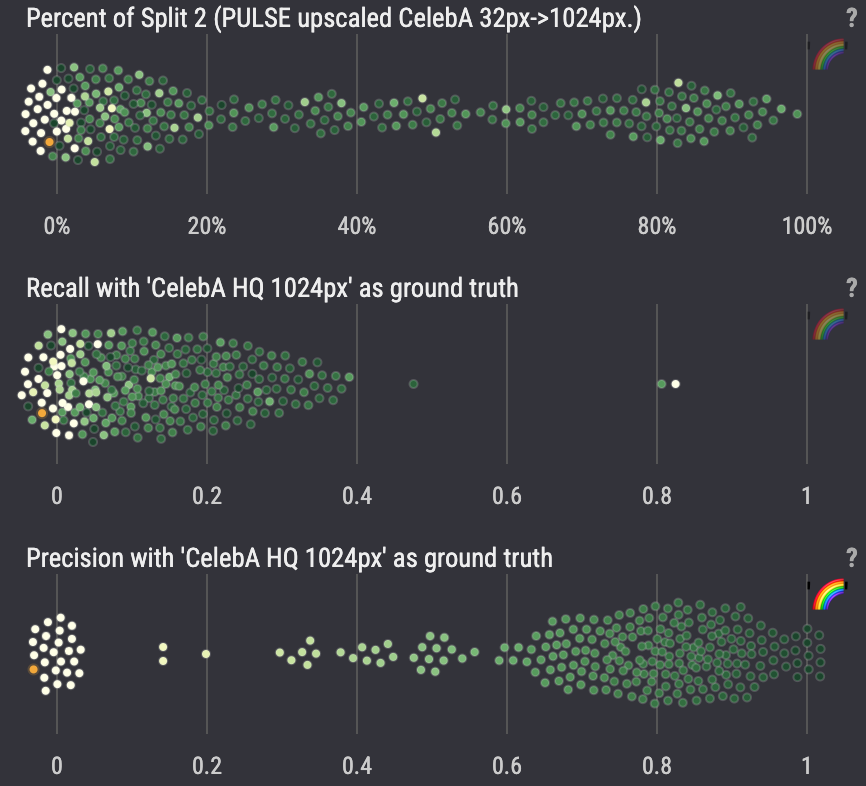}
    \caption{Cluster metric charts colored by cluster precision. On the left of the precision chart is a group of clusters with very low precision that are mostly composed of images only from the ground truth data, many of these also have low recall, suggesting the algorithm struggles to produce any output for them.}
    \label{fig:celeb_a_by_precision}
\end{figure}

\begin{figure}[h]
    \centering
    \includegraphics[height=7.5cm]{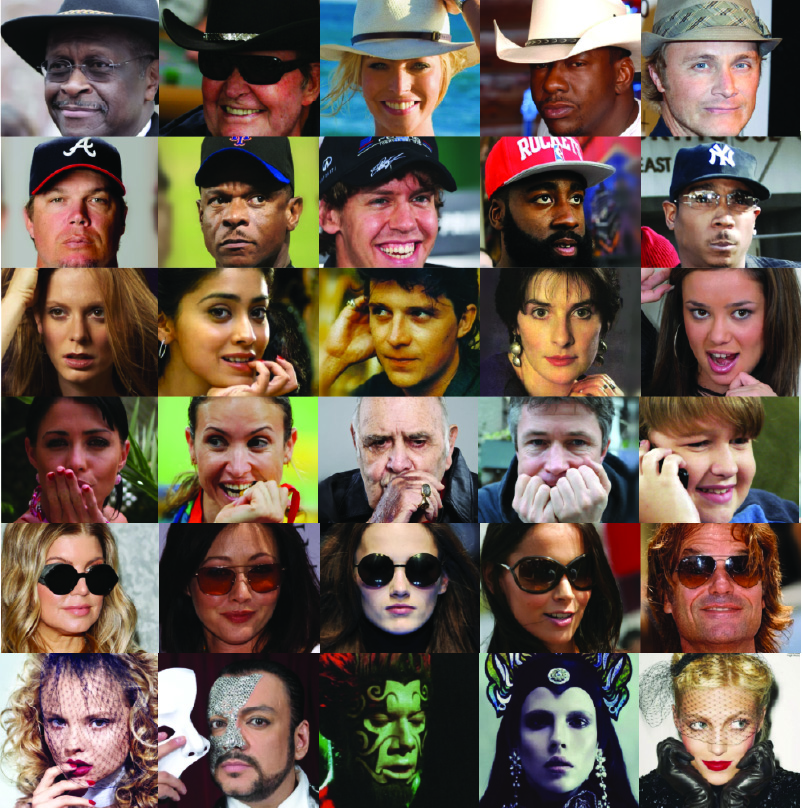}
    \caption{Each row contains samples from a different cluster with that only has images from the ground truth data and none from the algorithm output. These are examples of images the algorithm struggles to upscale.}
    \label{fig:celeb_a_zero_precision}
\end{figure}

Our analyst sets the number of clusters to 250 and the original CelabA-HQ images on the left split, and immediately notices a group of clusters on the lowest end of the precision chart. She notices that many clusters also have low recall and are mostly \textit{composed of images only from the ground truth} (see \autoref{fig:celeb_a_by_precision}). She clicks on some low-recall/low-precision clusters and observes a few categories where the algorithm struggles to produce any output. These include: people wearing hats or other headgear, images where the person has a a microphone or hand in front of them, faces with a lot of facepaint or makeup, and images where people are wearing sunglasses. Each of these clusters has between 60-120 images and a few samples are shown in \autoref{fig:celeb_a_zero_precision}.

\emph{\textbf{What kinds of images does PULSE struggle to produce high quality output for?}}

\begin{figure}[h]
    \centering
    \includegraphics[width=\linewidth]{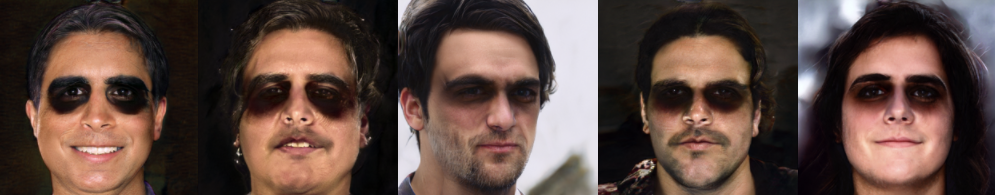}
    \caption{Samples from a cluster of upscaled images of people wearing sunglasses.}
    \label{fig:celeb_a_sunglasses}
\end{figure}

She continues to explore low precision clusters by scanning left to right along the chart. One of the lowest precision clusters contains ground truth images of people wearing spectacles (and a few wearing sunglasses), however from the algorithm output she sees a number of low quality (i.e. unrealistic) images that are likely up-scaled from ones where the subject is wearing sunglasses (\autoref{fig:celeb_a_sunglasses}). She refines her hypothesis about what the algorithm does to portraits with sunglasses, determining that, \textit{"if a person is wearing sunglasses, the algorithm often fails to produce any image, and when it does, it produces unrealistic output"}.

\emph{\textbf{What kinds of images does PULSE do well at upsampling?}}

The precision chart can also be used to look at what images PULSE does well at. As our analyst browses clusters on the right side of the distribution, she observes that a majority of high quality outputs seem to be of lighter skin tone faces that look relatively young or middle aged. 

Her manual inspection reveals some of the model's failure modes and strengths along demographic categories even though she does not have access to quantitative measures of performance across sensitive features such as skin tone or age.

\section{Domain Expert User Study}

We evaluated Ravel in a two-stage study, where the first stage identified domain experts' model evaluation goals, workflows and existing tools, and the second stage investigated the usability and utility of the Ravel UI. Participants were recruited from a convenience sample of full time employees as a large technology company, and there was a diversity in gender identity ($n_{female}$ = 1, $n_{Male}$ = 5), product area (5 different teams), and office location ($n_{Israel}$ = 2, $n_{United States}$ = 4). Stage one (n = 4) included four expert research scientists and software engineers who currently work on training and evaluating generative models, and stage two (\textit{n} = 5) included three of the users from stage one plus two additional users who have experience working with generative models. Both evaluation stages used remote moderated Google Meet video calls to speak with users and allow them to share their screen to show how they interacted with Ravel.

\subsection{Stage One: The Current State of Evaluation} \label{current state of eval}

Stage one aimed to understand the current state of evaluation for generative image model output. Semi-structured interviews were used, with questions about participants' familiarity and day to day work with generative models, goals and motivation for model evaluation, as well as current workflows and practices. All four users currently work on generative models, and were familiar with StyleGAN and its variants, BigGAN and its variants, as well as other models trained on ImageNet and FFHQ. There was diversity in architectures they work with and the tasks they apply generative models to. Here, we report the most common practices among users.

\subsubsection{Evaluation Goals \& Workflows}
All four participants reported that publication of model performance/improvement was a primary goal for their work. All four participants also expressed that evaluating output image quality was critical to understanding model performance.

All four participants reported a mix of quantitative and qualitative evaluation methods. In general, their workflows could be characterized by a common pattern of: model training accompanied by limited visual inspection for sanity-checking $\rightarrow$ examination of metrics $\rightarrow$ continued training $\rightarrow$ reexamination of metrics until a predetermined threshold is reached $\rightarrow$ qualitative visual inspection of image output. 

Two participants indicated that they believed the current “best practice” for determining image quality was human qualitative evaluation, often from crowdsourced studies. Importantly, all four participants also emphasized the ubiquitous reliance upon and limitations of FID scores in model evaluation. While one user suggested that “FID scores are much better than inception scores and other metrics”, they also conceded that “there is not a gold standard for evaluation of generated images”. All users indicated that they were dissatisfied with FID score as a primary evaluation metric, and that they had all experienced and read cases where they felt that FID score did not align with human inference, a sentiment supported in Zhou, et al.  \cite{zhouHYPEBenchmarkHuman}. 

\subsubsection{Evaluation Tools}
All four users mentioned using TensorBoard \cite{abadiTensorFlowLargeScaleMachine2016a} or Colab\footnote{https://colab.research.google.com/} (a computational notebook envrionment similar to Jupyter Notebook \cite{Kluyver2016jupyter}) to examine model output. For Colab, the main strength reported was flexibility. The flexibility of Colab enables bespoke analyses that we elaborate on in section \ref{expert_feedback}. Its main limitations were difficulties reusing code between projects and sharing results, especially with non-technical collaborators or stakeholders not directly involved with model training. For TensorBoard, the main strengths reported were ease of use during training to quickly visualize metrics and see sample output at different stages of training. The main limitations mentioned about TensorBoard were its latency when loading many images, and lack of customization compared to an open ended tool like Colab.

\subsubsection{Qualitative Visual Evaluation Tasks}
Determining image quality was reported as the most important task. Determining the diversity of images was also important to all four users, and one user mentioned that looking for the occurrence of mode collapse was an explicit part of their evaluation workflow. All users indicated that it was important to do more granular and bespoke image generation, including looking at samples within certain classes or samples close to each other in embedding space. One user discussed examining different levels of truncation for latent vectors to make decisions about both realism and aesthetics in the output. Users reported viewing a small number of images. One user reported looking at approximately 64 images, and never more than 100 images, per evaluation step of the model training pipeline. Another user explained that when they work with a team on an evaluation, they typically generate and inspect about 50 images. However, when working alone, this user said that they would inspect at least 200 images, noting that it was difficult to get a group consensus on more than 50 images at a time. 

We determined four categories of critical evaluation tasks from this part of the study: image diversity, image quality, mode collapse, and a “catch-all” category of additional explorations of classes and samples.

\subsection{Stage Two: Observing how Ravel is Used}
The objective of this stage was to determine how the Ravel UI can be used for the critical evaluation tasks identified in stage one. A brief slide deck and video explaining the UI components was sent to users upon confirmation of their participation, which they were asked to review before the study session. The study sessions themselves lasted for one hour and consisted of task based user exploration of the tool and semi-structured interview questions.

\subsubsection{Task one: Diversity}
The first task we asked users to attempt was to decide whether BigGAN or BigGAN-deep was performing better in terms of the diversity of sea urchin images, in a setup mirroring the Ravel instance described in Section \ref{sea urchins}. This instance showed 128x128 pixel images from both models, with 50,000 images per model and 50 images per class from each model. The UI resolution was set to show the same number of images for each user: 5 images per row in each split.  

Users were presented with an instance of Ravel with the "sea urchin" class selected in the \textit{Highlighted Classes} menu, showing results from BigGAN and BigGAN-deep in the left split and right split, respectively. On initial load the instance displayed the cluster containing most of sea urchin images selected: 39  (out of 50) urchins from BigGAN and 45 (out of 50) urchins from BigGAN-deep. 

All users started by attending to images in each split. No users examined metrics or asked about metrics as a first step. This emphasizes the importance of Ravel intuitively depicting some of the most important information for assessing image diversity: a salient grid of sample images within a cluster of interest. Three users immediately noted the number of samples in each split, paying close attention to which model had a greater number of samples in the cluster of interest. The most common flow involved inspecting individual images and counting or “eyeballing” the number of images in each model with unique background colors, sea urchin poses, and sea urchin colors (all users mentioned these three features). Three users scanned the metrics charts and clicked on the other highlighted clusters with sea urchins. All five users made the determination that BigGAN-deep was performing better in terms of the diversity of sea urchin images, with one user explicitly stating that this confirmed their prior expectations. This demonstrates consistency in how users complete this critical task with Ravel, and shows the potential for convergent decision making and operationalization of workflows in qualitative visual evaluation. It is also notable that Ravel helped confirm one user's prior expectations about BigGAN-deep’s diversity performance, although that could also have been a potential biasing factor in their evaluation process.

\subsubsection{Task two: Quality}
For the second task, users were asked to decide whether BigGAN or BigGAN-deep was performing better in terms of the quality of golden retriever images. Users were presented with the same Ravel instance as task one, but this time the "golden retriever" class was selected in the Highlighted Classes menu and a cluster showing most of the golden retriever images was selected: 46 (out of 50) in the BigGAN split and 46 (out of 50) in the BigGAN-deep split.

All users immediately noted many artifacts in output images from both models. All users clicked on individual images and narrated particular issues, such as the shape of dogs’ noses, the number of legs, and the accuracy of the form and pose of the dogs in each sample. The most common workflow was to count or “eyeball” the number of artifacts in each split to make an initial determination, but evaluation workflows were notably more diverse between users for the rest of the task. Five users reported that this task was more difficult than the diversity task while one user reported that it was easier. Some users indicated that FID and Inception score for each model would be an important part of making this determination in their typical workflow. Four out of the five users made the determination that BigGAN-deep also performed better in terms of the quality of golden retriever images, and one user did not make an explicit determination for this task. Once again, Ravel supported consistency in the primary image quality evaluation strategy across all users, but individual exploration varied after this initial decision. Some users validated their visual judgments by looking at recall and precision charts, but others simply explored the charts without forming additional opinions about the task.

\subsubsection{Task three: Mode Collapse}
Following the Quality task, users were asked if they were familiar with the term mode collapse and if it was something they used to evaluate generative model output. Four out of five users were familiar with mode collapse and reported it as a useful discovery in the model evaluation process, while one user was unfamiliar with the term. Users who were familiar with mode collapse were then asked to use Ravel to determine whether it had occurred for either model within any class.

Because this task was not constrained to a single class, it was more open-ended, and thus more challenging for users to make a determination about. Workflows varied widely between users, as they were given no guidance about how to accomplish the task or make a determination. Several users mentioned that they expected BigGAN to exhibit more mode collapse after observing that BigGAN-deep had better diversity in the first task. Four users who found concrete instances of mode collapse (e.g. Fig. \ref{fig:stylegan2-facepaint-artefacts}) did so by selecting clusters with the smallest values in the Recall chart. Four out of five users viewed the cluster samples displayed in figure \ref{fig:BigGAN-deep_spacebar_typewriter} at some point during their exploration, but one user did not identify it as mode collapse, and one user did not select it at all. Nevertheless, this was a promising observation, and demonstrates further consistency of Ravel’s usability and specific utility of the recall visualization. There was majority agreement that the recall chart can be used to identify mode collapse, and it was revealed to occur more often in BigGAN-deep.

\subsubsection{Task four: Additional exploration of classes, samples, and features}
For their final task, users were asked to view an instance of Ravel showing generated images from a StyleGAN2-based model in the right split, and images from its ground-truth training data, FFHQ, in the left split. Users were told they could freely explore the interface, either repeating the quality and diversity assessments or trying a new task that they might be interested in. 

This task was fully unguided, but most users started by assessing image quality for the StyleGAN2-based model. It was common for users to select clusters at the low and high ends of the Precision distribution. One user observed that clicking on a cluster with high precision revealed “typical FFHQ images…very good images without occlusion, faces looking at the camera, showing people with straight hair, and the model output is very similar to these images.”. Four out of five users discovered and explicitly verbalized that StyleGAN2-based model struggled to generate images of people with facepaint. They did this by clicking on clusters with low precision and noticing artifacts on many of the generated images. All four of these users expressed surprise upon making this new discovery about the model. This demonstrates that even without a specific prompt, many users will make the same kinds of discoveries and follow the same types of evaluation processes with Ravel. 
One of our participants was on the team that had originally trained the StyleGAN2-based model that we used and discovered that the model was unable to generate faces of people wearing a particular style of fuzzy winter hat that was fairly common in the ground truth data. He remarked that he wasn't aware of that inability and was pleasantly surprised to be able to discover it.

The other common tasks were “semantic explorations” of the clusters and the embedding space. Four out of five users examined images in the FFHQ split to make decisions about whether the clusters were semantically meaningful, and to explore the diversity of FFHQ. Two users pondered whether or not the embedding space was doing a good job of capturing what they, as humans, would group together. These users investigated the proximity of samples in the Samples UMAP chart and determined that the ImageNet feature space is not optimal for clustering faces, since they could not find consistent visual similarity between nearby samples in some of the clusters. 

\subsubsection{Discussion of Expert Feedback} \label{expert_feedback}
Users were overall impressed with the tool. All five users thought that Ravel would fit into their current evaluation process, and that it was especially useful for researchers who publish generative models. Here, we summarize additional feedback from the reflection portion of stage two.

Two users reported that their exploration made them doubt whether the Inceptionv3 features were good for evaluating face portrait generation. In reflection, one of these users stated that Ravel could be used to learn about the feature space itself, which could be broadly useful in generative image model evaluation. 

Upon discovering mode collapse in BigGAN-deep, one user stated that Ravel would be useful for probing state-of-the-art models to learn which classes they can’t generate images for, which could point to systematic failure modes for researchers to focus on improving. 

One user noted that Ravel could help their team reach conclusions about model performance \textit{more quickly than using Colab}, especially for understanding the diversity of images. They explained that using Colab required developing a bespoke visualization tool and manual calculations of FID in each cluster, whereas the same type of useful information was readily available in the Ravel UI. Two users emphasized that Ravel could help operationalize how researchers evaluate quality and diversity, one of which explained that Ravel could be “a forcing function for having a standard UI/pipeline for results”, arguing that this would make it “easier to share results with someone on another team, or someone non-technical…even with my manager who doesn’t have time to run my code”. This confirms for us that there is a place for bespoke purpose built interfaces like Ravel, that while less flexible than Colab, are more purpose built for common evaluation tasks that researchers perform.

Overall, the expert feedback from stage two was positive and enthusiastic, with several users expressing excitement about continued exploration in Ravel and incorporating it into their own workflows.

\section{Limitations and Future Work}
Our qualitative user study was performed with a small, domain expert sample from a single company, and therefore may not be an externally valid representation of all researcher experiences with generative image models. The study sessions were also limited to one hour per user, with each evaluation task time-boxed to 15 minutes or less. Richer and more diverse interactions and discoveries could be possible with a longer duration of tool use.

Two users wanted to see images at a higher resolution, and two other users wanted to see the original resolution of the images when examining them for artifacts; this is not an inherent issue with Ravel but is an important design affordance for the future. Two users wanted to be able to mark images in each split once they had viewed them, and enabling this would support the 'counting' based workflows we saw participants use to complete the tasks.

When using the class conditional model, users initially reported that they would prefer to see all images from a given class in the same view (i.e. clustering by class label) to make a judgement about that class. However on further exploration they noted it was useful to see 'outlier' images for a given label in context with the other images they are most similar to. This suggests that both workflows are important to support for class conditioned models and should be supported by tools like Ravel.

The authors also note a number of limitations of the system we observed while watching users use the tool:

User's cannot always interpret the 'meaning' of clusters (i.e., construct a rationale for why a set of images are clustered together). This is a general issue with unsupervised methods like clustering, but we found that users would often try to attach some semantically meaningful description to each cluster to ground their comparison.

Exploring ways to 'describe' clusters or summarize differences between clusters could be important future work to aid user comprehension. We also think exploring other clustering methods, in particular hierarchical methods, could be a particularly attractive means to produce clusters at different levels of granularity to help build understanding of groups within the dataset.

In describing the Sample viewer (section \ref{sample viewer}) we noted that one limitation is that not all of the images are visible in one screen if there are many images present in the cluster. While one mitigation is to decrease the size of clusters by increasing the number of clusters, we believe that future work could provide better ways to get the visual gestalt of the entire cluster in one view. Possibly adapting techniques such as those described in Activation Atlas \cite{carter_activation_2019}, creating stacks of very similar images within a cluster, or other ways of sub-sampling or sorting to ensure that we are displaying the maximum variety about a cluster in a single screen.

One user task that Ravel does not directly support is detecting memorization. One user commented that adding a real time nearest-neighbor search to the interface would likely make it useful for this task.

\section{Conclusion}

We presented Ravel, a visual analysis tool that enables researchers to perform large scale qualitative evaluation of generative model outputs.
Our primary contributions included:
\begin{itemize}
    \item A visual analytics system that supports \textit{multiple evaluation tasks} (e.g. evaluating quality \& diversity, discovering mode collapse or gaps in model output) for generative image models and is agnostic to model architecture and internals.
    \item \textit{Interactive exploration of large generative image model datasets}, facilitated by clustering and the use of fine grained visualization of cluster metrics to guide qualitative evaluation.
    \item A user interface that uses \textit{visual comparison} driven by semantically meaningful embedding spaces to support reasoning about differences between image distributions and generate hypotheses about model behaviour.
\end{itemize}

The expert users in our study were able to generate consistent insights about model behaviour including identifying areas of the \textit{true data distribution the model was not capturing}, such as face paint or certain kinds of headgear in the StyleGAN2-based model or mode collapse in BigGAN-deep. This kind of insight is an example of one that is not possible to get from just looking at quantitative metrics like FID.

Our study participants confirmed our hypotheses that single number metrics are not fully sufficient measures of model performance. In addition to exploring metrics at greater granularity, Ravel allows users to explore metrics at finer granularity, revealing areas of model output where metrics like recall or precision do not capture problems in the generated images. Our users hypothesized that the underlying InceptionV3 embedding, used both in our tool and in the primarily metrics in the field, may not attend to certain kinds visual artefacts that are easily visible to humans. We believe that future work in this direction could enable better understanding of limits of the embedding spaces themselves and how they affect both metrics and the workflows that use them.

\ifCLASSOPTIONcompsoc
  \section*{Acknowledgments}
\else
  \section*{Acknowledgment}
\fi

The authors wish to thank Mario Lucic, Marvin Ritter, Ben Poole, Han Zhang, Chitwan Saharia, James Wexler and Lucas Dixon for their help and feedback on this work. We also thank our study participants for their feedback.



\bibliographystyle{IEEEtran}
\bibliography{paper}

\end{document}